\def\year{2023}\relax
\documentclass[letterpaper]{article} 
\usepackage{aaai22}  
\usepackage{times}  
\usepackage{helvet}  
\usepackage{courier}  
\usepackage[hyphens]{url}  
\usepackage{graphicx} 
\usepackage{hyperref}
\usepackage{makecell}
\newcommand{\term}[1]{\begin{small}\texttt{#1}\end{small}}
\usepackage{natbib}  
\usepackage{caption} 
\DeclareCaptionStyle{ruled}{labelfont=normalfont,labelsep=colon,strut=off} 
\usepackage{algorithm}
\usepackage{algorithmic}
\usepackage{xcolor,,colortbl}
\usepackage{multirow}
\usepackage{arydshln}

\usepackage{listings}

\colorlet{punct}{red!60!black}
\definecolor{background}{HTML}{EEEEEE}
\definecolor{delim}{RGB}{20,105,176}
\colorlet{numb}{magenta!60!black}

\lstdefinelanguage{json}{
    basicstyle=\footnotesize\ttfamily,
    numbers=left,
    numberstyle=\scriptsize,
    stepnumber=1,
    numbersep=8pt,
    showstringspaces=false,
    breaklines=true,
    frame=lines,
    backgroundcolor=\color{background},
    literate=
     *{0}{{{\color{numb}0}}}{1}
      {1}{{{\color{numb}1}}}{1}
      {2}{{{\color{numb}2}}}{1}
      {3}{{{\color{numb}3}}}{1}
      {4}{{{\color{numb}4}}}{1}
      {5}{{{\color{numb}5}}}{1}
      {6}{{{\color{numb}6}}}{1}
      {7}{{{\color{numb}7}}}{1}
      {8}{{{\color{numb}8}}}{1}
      {9}{{{\color{numb}9}}}{1}
      {:}{{{\color{punct}{:}}}}{1}
      {,}{{{\color{punct}{,}}}}{1}
      {\{}{{{\color{delim}{\{}}}}{1}
      {\}}{{{\color{delim}{\}}}}}{1}
      {[}{{{\color{delim}{[}}}}{1}
      {]}{{{\color{delim}{]}}}}{1},
}

\usepackage{newfloat}
\usepackage{listings}
\floatstyle{ruled}
\newfloat{listing}{tb}{lst}{}
\floatname{listing}{Listing}
\title{Wiki-based Communities of Interest: Demographics and Outliers}
\author{
    Hiba Arnaout\textsuperscript{\rm 1}
    Simon Razniewski\textsuperscript{\rm 1},
    Jeff Z. Pan\textsuperscript{\rm 2}
}
\affiliations{
    \textsuperscript{\rm 1}Max Planck Institute for Informatics, Saarbrücken, Germany\\
    \textsuperscript{\rm 2}The University of Edinburgh, Edinburgh, United Kingdom\\

}

\usepackage{bibentry}

\begin{document}

\maketitle

\begin{abstract}
In this paper, we release data about demographic information and outliers of communities of interest. Identified from Wiki-based sources, mainly Wikidata, the data covers 7.5k communities, such as members of the White House Coronavirus Task Force, and 345k subjects, e.g., Deborah Birx. We describe the statistical inference methodology adopted to mine such data. We release subject-centric and group-centric datasets in JSON format, as well as a browsing interface. Finally, we forsee three areas this research can have an impact on: in social sciences research, it provides a resource for demographic analyses; in web-scale collaborative encyclopedias, it serves as an edit recommender to fill knowledge gaps; and in web search, it offers lists of salient statements about queried subjects for higher user engagement. 
\end{abstract}

\section{Introduction}
\subsection{Motivation}
A community consists of a group of people who share a commonality such as geography (\textit{Texans}), religion (\textit{Christian}), ethnicity (\textit{Arab}), or a combination (\textit{Arab Texans}). One commonality that is less often discussed are communities of passion or purpose, the so-called \textit{communities of interest}~\cite{fischer2001communities}. This refers to groups of people who share a profession, practice, or an interest. For instance, members of \textit{White House Coronavirus Task Force} is a community of practitioners in the medical field. Not to be confused with the much broader community of \textit{all} medical practitioners, we focus on contextualized groups of people. In this case, people who were appointed by the \textit{White House} for a specific task. A second example is recipients of \textit{ACM Fellowship}, rather than \textit{all} computer scientists. This allows for fine-grained analyses over various topics, cultures, time frames, etc.

One standard task for understanding communities is identifying their demographic factors. Demographics are statistical information about a community that includes such factors as gender, occupation, linguistic background, nationality, and location~\cite{ashraf2020demographic}. In geo-based communities for example,  identifying demographics can contribute in local policy making or in understanding consumer behavior for national businesses. In communities of interest, it could contribute to identifying under-represented groups or understanding cultural differences between similar communities across countries or continents. Moreover, compiling top demographics facilitates the task of finding outliers, i.e.,  members that have different characteristics than the majority, e.g., \textit{Deborah Birx} is a female while 86\% of the \textit{White House Coronavirus Task Force} members are male. These can contribute in studies of under-represented groups in different settings.

While extensive research has been conducted on demographics of large geo-based communities~\cite{chambers_jonathan_2020_3768003,doi:10.1080/0032472031000149566} and topic-specific study cases~\cite{Poschmann2022,sun2021men,zhou2020prevalence}, to the best of our knowledge, this is the first work to address communities of interest,  releasing data which covers 16 topics from 4 main domains, namely \textit{culture}, \textit{geography}, \textit{history \& society}, and \textit{STEM}.

\begin{table}
\resizebox{0.46\textwidth}{!}{\begin{tabular}{c|c|c}
\textbf{\cellcolor{gray!15}Community} & \textbf{\cellcolor{gray!15}Demographics} & \textbf{\cellcolor{gray!15}Outliers}\\
\makecell{members of White House\\ Coronavirus Task Force} & \multicolumn{1}{c|}{\makecell{male\\republican}}& \makecell{Deborah Birx\\(female) \\ Jerome Adams\\(independent)}\\
\hdashline
 \makecell{members of Indian\\National Science Academy} & \multicolumn{1}{c|}{\makecell{male\\indian}} & \makecell{Jörg Hacker\\(german)}\\
 \hdashline
 \makecell{winners of ACM Fellowship}  & \multicolumn{1}{c|}{\makecell{male\\american\\computer science}} & \makecell{Susan Nycum\\(female, lawyer)\\Calvin Gotlieb\\(canadian)\\}\\
\end{tabular}}
\caption{Demographics and outliers of 3 communities.}
\label{tab:samples}
\end{table}

\subsection{Contributions}

We construct a dataset to capture demographics and outliers about communities of interest from  Wikidata~\cite{WD}. Table~\ref{tab:samples} shows three sample communities of interest, including winners of \textit{ACM Fellowship}. While most winners are \textit{male American computer scientists}, two outliers are \textit{Susan Nycum} who is \textit{not} male neither a computer scientist, but a female lawyer, and \textit{Calvin Gotlieb} who is not \textit{American} but \textit{Canadian}. In a nutshell: we collect communities of interest from Wikidata using pre-defined properties~\footnote{\href{https://www.wikidata.org/wiki/Wikidata:List_of_properties}{https://www.wikidata.org/wiki/Wikidata:List\_of\_properties}} indicating \textit{interest}, inspired by the idea for collecting peer groups in ~\cite{jws21arnaout}. For example \textit{position held} indicates people holding the same public office and \textit{member of} indicates people belonging to the same organization, club or group. We then query its topic and domain using the Wiki-topic tool. In the case of \textit{ACM Fellows}, the domain is \textit{STEM} and the topic is \textit{Computing}. Next, we query Wikidata for demographic statements about members of the community using demography-describing properties in Wikidata, e.g., \term{gender (P21)}. We compute the most frequent value for each of the factors, e.g., \term{occupation-computer scientist} (0.93) and \term{gender-male} (0.80). Finally, we use common factors to identify outliers in the community. These are members whose characteristics do not fully match the community's demography profile. For example, members who are \textit{not} computer scientists. This results in a group-centric dataset. In addition, we  construct a subject-centric dataset, by merging for every subject, the list of salient statements across multiple communities they are a member of.  

Our datasets can be downloaded from the Zenodo sharing service at: \href{https://doi.org/10.5281/zenodo.7410436}{https://doi.org/10.5281/zenodo.7410436}. We release the data in an easy to parse JSON format.  We also publish a web interface for friendly browsing: \href{https://wikiknowledge.onrender.com/demographics/}{https://wikiknowledge.onrender.com/demographics}.

On usage, we believe this dataset can provide useful seeds in social computing problems, by identifying demographic data for a better analysis of communities of interest and action-warranting under-representations, e.g., gender or ethnicity-bias in certain communities. Downstream applications include comparing similar communities within the same culture/country, e.g., the majority of winners of the \textit{Latin Grammy Award for Best New Artist} are men while \textit{BET Award for Best New Artist} are women (both American musical awards for newcomers). Moreover, the data can give insights into equivalent communities of interest between different cultures, e.g., the typical \textit{Prime Minister of Iran} is \textit{Shia} but the typical \textit{Prime Ministers of Lebanon} is \textit{Sunni}, which are the two main religious branches of \textit{Islam}. One might be intrigued to look more into the historical reasons of these findings or predict future election winners. Beside social sciences, this data can be useful for knowledge curation of collaborative knowledge bases such as Wikidata by providing the curators with edit recommendations about missing information, e.g., a missing profession of a member in a community can be derived from its common profession (the typical \textit{Turing Award} recipient is a  computer scientist). Finally, producing salient provenance-extended statements about outliers can increase user engagement in web search~\cite{funfacts}, e.g., \textit{unlike 560 out of 636 recorded winners of Presidential Medal of Freedom, Stephen Hawking is not an American, but British, recipient.}

\section{Dataset Creation}
\label{sec:creation}

\subsection{Identifying Communities}
We choose Wikidata as our source of communities of interest. It is a web-scale collaborative knowledge base which covers 97\% of Wikipedia articles (called items). For example, \textit{ACM Fellowship} as an article (\href{https://en.wikipedia.org/wiki/ACM\_Fellow}{https://en.wikipedia.org/wiki/ACM\_Fellow}) and   (\href{https://www.wikidata.org/wiki/Q18748039}{https://www.wikidata.org/wiki/Q18748039}) as an item. 

Wikidata, and structured stores of data in general, are a good source for constructing communities of interest, due to their well-canonicalized entities and properties, i.e., using IDs. For instance, one does not have to worry if an \textit{Oscar} is referred to using the word \textit{Oscar} or \textit{Academy Award} or different wording but can simply query it using its unique ID \term{Q19020}. Moreover, Wikidata contains additional information about every item, including temporal signals and links to Wikipedia articles, which allows for more sophisticated use cases, e.g., demographics in a certain era. We pick 6 properties indicating interest or profession (\term{position held} - \term{P39}, \term{award} - \term{P166}, \term{participant in} - \term{P1344}, \term{candidate in election} - \term{P3602}, \term{nominated for} - \term{P4353}, \term{member of} - \term{P463}). We instantiate a SPARQL query~\footnote{\href{https://query.wikidata.org/}{https://query.wikidata.org/}} with a property of interest and one of its objects (community) to collect its members, e.g., \term{select distinct ?subject where \{?subject wdt:P166 wd:Q18748039\}} is used to collect \textit{ACM Fellowship} recipients. A list of subjects is returned, including \textit{Thomas Henzinger}, \textit{Susan Nycum}, \textit{Calvin Gotlieb}, etc. 

The outcome of this step is 7.5k communities of interest covering 16 topics and 345k subjects. Given a community of interest, the Wiki-topic tool~\footnote{\url{https://wiki-topic.toolforge.org/topic}} is queried for top-3 topics. An overview is shown in Table~\ref{tab:topicoverview}~\footnote{For readability, we only display topics with $\geq$ 50 communities.}. Note that a community can belong to more than 1 topic, e.g., \textit{Presidents of the Senate of Nigeria} is both related to \textit{Politics} under \textit{History \& Society} and to \textit{Africa} under \textit{Geography}.

\begin{table*}
\centering
\resizebox{0.9\textwidth}{!}{\begin{tabular}{l|c|c|l}
\textbf{\cellcolor{gray!15}Topic} & \textbf{\cellcolor{gray!15}Domain} & \textbf{\cellcolor{gray!15}\# of Communities} & \textbf{\cellcolor{gray!15}Sample Community (\textit{title}, \textit{recorded members}, \textit{sample member})}\\
Biography & \multirow{7}{*}{\rotatebox[origin=c]{90}{Culture}} & 1358 & \textit{Winners of Suffrage Science award}, \textit{35}, \textit{Pippa Goldschmidt}\\
Literature &   & 416& \textit{Winners of EU Prize for Literature}, \textit{96}, \textit{Magdalena Parys}\\
Media &  &  1071& \textit{Winners of Academy Award for Best Actress}, \textit{79}, \textit{Emma Stone}\\
Performing Arts &  & 74 & \textit{Winners of Special Tony Award}, \textit{21}, \textit{Judy Garland}\\
Philosophy \& Religion &  &  250 & \textit{Popes}, \textit{269}, \textit{John Paul II}\\
Sports &   & 267 & \textit{Winners of NBA Coach of the Year Award}, \textit{52}, \textit{George Karl}\\
Visual Arts &  & 189 & \textit{Members of Royal Academy of Arts}, \textit{840}, \textit{William Etty}\\
\hdashline
Africa & \multirow{5}{*}{\rotatebox[origin=c]{90}{Geography}} & 191 & \textit{Presidents of Uganda}, \textit{13}, \textit{Idi Amin}\\
Americas &  & 927 & \textit{Governors of Wisconsin}, \textit{47}, \textit{Jim Doyle}\\
Asia &  & 947& \textit{Chiefs of the Philippine National Police}, \textit{23}, \textit{Rodolfo Azurin Jr.}\\
Europe &  & 2737& \textit{Prime Ministers of Poland}, \textit{44}, \textit{Jan Olszewski}\\
Oceania &  & 247& \textit{Winners of New Zealand Order of Merit}, \textit{148}, \textit{Charles Higham}\\
\hdashline
Business \& Economics & \multirow{6}{*}{\rotatebox[origin=c]{90}{History}} & 148& \textit{Winners of Queen Elizabeth Prize for Engineering}, \textit{11}, \textit{Tim Berners-Lee}\\
Education &  & 134& \textit{Chancellors of the University of Oxford}, \textit{28}, \textit{Roy Jenkins}\\
History &  & 252& \textit{Dukes of Normandy}, \textit{26}, \textit{William the Conqueror}\\
Military \& Warfare & & 386& \textit{United States secretaries of War}, \textit{49}, \textit{William Howard Taft}\\
Politics \& Government &  & 950 & \textit{Leaders of the House of Commons}, \textit{80}, \textit{Andrew Lansley}\\
Society & & 63 & \textit{Winners of Civil Courage Prize}, \textit{36}, \textit{Alexei Navalny}\\
\hdashline
Biology & \multirow{6}{*}{\rotatebox[origin=c]{90}{STEM}} & 116 & \textit{Winners of Darwin Medal}, \textit{67}, \textit{Ernst Haeckel}\\
Chemistry &  & 63 & \textit{Winners of ACS Award in Pure Chemistry}, \textit{92}, \textit{Stuart Schreiber}\\
Earth \& Environment &  & 76& \textit{Winners of Paleontological Society Medal}, \textit{54}, \textit{Alfred Romer}\\
Mathematics &  & 75& \textit{Winners of SIAM Fellow}, \textit{537}, \textit{Ernst Hairer}\\
Medicine \& Health &  & 98& \textit{Winners of Pollin Prize for Pediatric Research}, \textit{17}, \textit{Basil Hetzel}\\
Physics &  & 120& \textit{Winners of Nobel Prize in Physics}, \textit{222}, \textit{Albert Einstein}\\
Space &  & 55 & \textit{Winners of NASA Distinguished Service Medal}, \textit{319}, \textit{Frank Borman}\\
\end{tabular}}
\caption{Overview of covered topics and sample communities.}
\label{tab:topicoverview}
\end{table*}

\subsection{Defining Demographic Factors}
Now that we have communities of interest with their members and topics, we want to identify their most frequent values given a set of standard demographic factors~\cite{ashraf2020demographic}. We map each of those to equivalent Wikidata properties (see Table~\ref{tab:demographicfactors}). For instance, we identify the nationality of a certain member using property \term{P27}.

\subsection{Inferring Demographics and Outliers}
At this point, we have all the ingredients to start collecting community demographics and outliers. For every community, e.g., \term{award-ACM Fellow}:
\begin{enumerate}
    \item From Wikidata, query values for the predefined demography-properties, e.g., \term{gender}(\textit{Thomas Henzinger}, \textit{Calvin Gotlieb}, ..) = \term{male}.
    \item Compute relative incidence of each factor-value pair, e.g., \# male recipients of \textit{ACM Fellowship}/\# recipients of \textit{ACM Fellowship}= 673/839 = 0.80
    \item Sort by descending order of relative incidence, e.g., \term{occupation-computer scientist} (0.93),  \term{gender-male} (0.80), \term{nationality-U.S.} (0.58), etc..
    \item Collect outliers as members with demographic data not matching that of the top-k of the community, e.g., NOT(\term{gender-male}) applies to \textit{Susan Nycum} and NOT(\term{nationality-U.S.}) applies to \textit{Calvin Gotlieb}.
\end{enumerate}

\noindent
\textbf{Accuracy of Inferred Information.} When inferring non-asserted factors for certain members, one unavoidable challenge is the \textit{correctness} of these inferences. Due to the open-world assumption knowledge bases such as Wikidata postulates, absent statements can be either false (negative), or true but simply absent (missing positive). Present statements in some cases can also be undetectable using exact-match querying due to potential modelling issues. We remedy these using three heuristics: (i) \textbf{The partial completeness assumption PCA}~\cite{amie,dong2014data}, which asserts that if a subject has \textit{at least one} object for a given property, then there are no other objects beyond those that are in the knowledge base, e.g., if we have at least 1 award for subject \term{X} then we assume that their list of awards is complete. (ii) \textbf{Hierarchical checks}, where we exploit the type system, i.e., class taxonomy, in search for a contradiction of a certain negated factor. For instance,  \term{occupation-Catholic priest} does \textit{not} hold for subject \term{X}, but \term{occupation-Latin Catholic priest} does, and \term{(Latin Catholic priest, subclassOf, Catholic priest)} is a statement in Wikidata. Hence, if \term{X} is a \textit{Latin Catholic priest}, they are also a \textit{Catholic priest}. (iii) \textbf{Semantic similarity checks} to avoid possible synonymous or near-synonymous contradictions, we compute the sentence similarity between a candidate statement and an existing statement for subject \term{X}. We do so using SBert~\cite{reimers-2019-sentence-bert} with 0.6 as a similarity threshold, e.g., similarity (``\textit{teacher}'', ``\textit{professor}'') = 0.62, avoiding the inference that someone is a \textit{professor} but not a \textit{teacher} and vice versa. 95\% of the eliminated candidates are due to PCA, 2\% due to hierarchical checks, and due to 3\% semantic similarity checks.

\begin{table}
\centering
\begin{tabular}{l|l}
\textbf{\cellcolor{gray!15}Factor} & \textbf{\cellcolor{gray!15}Wikidata Property}\\
Gender	& sex or gender (\term{P21})\\
Sexual orientation	& sexual orientation (\term{P91})\\
Occupation	& occupation (\term{P106}) \\
Political leaning	& member of political party (\term{P102})\\
Religion	& religion or worldview (\term{P140})\\
Linguistic background	& native language (\term{P103})\\
Ethnicity \& race &	ethnic group (\term{P172})\\
Nationality &	country of citizenship (\term{P27})\\
Location	& residence (\term{P551})\\
\end{tabular}
\caption{Standard demographic factors and their properties.}
\label{tab:demographicfactors}
\end{table}

\section{Dataset Description}
We release two datasets on Zenodo~\footnote{\href{https://doi.org/10.5281/zenodo.7410436}{https://doi.org/10.5281/zenodo.7410436}} and a browsing interface~\footnote{\href{https://wikiknowledge.onrender.com/demographics/}{https://wikiknowledge.onrender.com/demographics/}}.

\subsection{Group-centric Dataset}
This dataset consists of 7530 rows in English language, with a total size of 64MB in JSON format. 
The fields of a JSON record are:
\begin{itemize}
    \item Title ID: title of the community using Wikidata IDs.
    \item Title label: title of the community using equivalent Wikidata labels.
    \item Number of recorded members: number of subjects in Wikidata that belong to the community. 
    \item Topics: a list of topics describing the community, e.g., \term{Culture.Media.Music}.
    \item Demographic factors: a list of top demographics, each consisting of an ID, value, and a score. The ID describes the factor using Wikidata identifiers and value has their labels. The score is the relative incidence within the community.
    \item Outliers:
    \begin{itemize}
        \item Reason: a statement on why the following members are considered outliers.
        \item Score: a numerical value indicating the frequency of this factor in the community.
        \item Members: a list of members for which this factor does not hold.
    \end{itemize}
\end{itemize}

A sample record~\footnote{For readability we omit some of the listed fields.} from the group-centric dataset:\\

\begin{lstlisting}[language=json,firstnumber=1,caption={JSON record from Group-centric dataset.},captionpos=b\label{lst:json1}]
{
	"title": "holders of position Lord Mayor of Dublin", 
	"recorded_members": 91, 
	"topics": ["Geography.Northern_Europe"], 
	"demographics": [
            "gender-male", 
            "occupation-politician"
	], 
	"outliers": [
		{
			"reason": "NOT(male) unlike 81 out of 91 recorded members", 
			"members": [
            "Catherine Byrne (female)",
            "Emer Costello (female)",
            "Alison Gilliland (female)"
            ]
		}, 
		{
			"reason": "NOT(politician) unlike 47 out of 91 recorded members", 
			"members": [
            "John D'Arcy (businessperson)",
            "Dermot Lacey (environmentalist)"
            ]
		}
	]
}
\end{lstlisting}

\subsection{Subject-centric Dataset}
We rerun the method described in the previous section but with two adjustments: for a given subject, we merge all outlier statements across different communities and rank them by descending order of incidence. Moreover, we extend the list of demography-properties to \textit{all} possible Wikidata properties. This subject-centric dataset consists of 345435 rows in English language, with a total size of 172MB in JSON format. 

The fields of a JSON record are:
\begin{itemize}
    \item Subject ID: the Wikidata ID of the subject.
    \item Subject label: its equivalent label.
    \item Statements: a list of salient statements across all communities of interest this individual is a member of:
    \begin{itemize}
        \item Statement ID: a statement using Wikidata ids.
        \item Statement label: a statement using Wikidata labels.
        \item Score: relative incidence.
    \end{itemize}
\end{itemize}

A sample record~\footnote{For readability we omit some of the described fields.} from the subject-centric dataset:\\

\begin{lstlisting}[language=json,firstnumber=1,caption={JSON record from the Subject-centric dataset.},captionpos=b\label{lst:json1}]
{
	"subject": "Serena Williams", 
	"statements": [
  	{
			"statement": "NOT(gender-male) but (female) 56 out of 68 recorded winners of L'Equipe Champion of Champions.",
			"score": 0.82
		},
  		{
			"statement": "NOT(sport-basketball) but (tennis) unlike 4 out of 8 recorded winners of Best Female Athlete ESPY Award.", 
			"score": 0.50
		}
	]
}
\end{lstlisting}

\subsection{Browsing Interface}
\begin{figure*}
\centering
\includegraphics[width=0.9\textwidth]{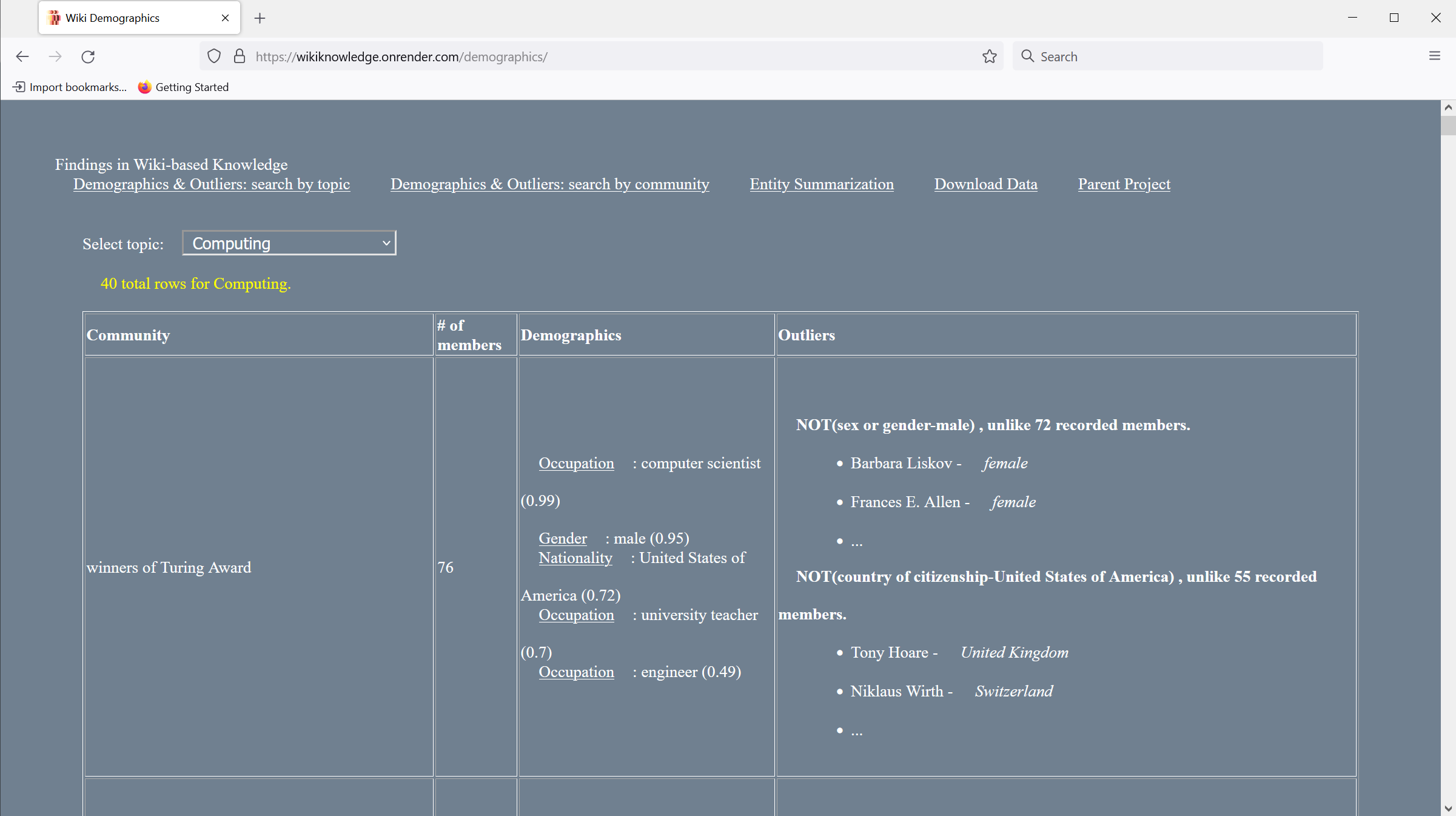}
\caption{Screenshot of the web interface, searching for demographics and outliers of communities under \textit{Computing}.} \label{fig:screenshot}
\end{figure*}

For users who wish to sample rows from each topic, we publish a web interface that can be accessed at: \href{https://wikiknowledge.onrender.com/demographics/}{https://wikiknowledge.onrender.com/demographics/}. A screenshot is shown in Figure~\ref{fig:screenshot}. It shows that the user selected \textit{Computing} as their topic of interest. It contains 40 communities about \textit{computing}, one of which is \textit{Winners of Turing Award} with top demographic factors as \textit{male computer scientists from the U.S.}. Outliers include women, e.g., \textit{Barbara Liskov} and non-\textit{Americans}, e.g., \textit{Tony Hoare} who is from the \textit{U.K.}. Users can also directly search for communities of their choice. On top of this feature, the website offers an entity summarization interface where users can query for the favorite public figures and get a list of top salient statements about them. 

\section{Applications}

\subsection{Demographic Data Analysis for Social Sciences} 
\begin{figure*}[t]
\centering
\includegraphics[width=0.8\textwidth]{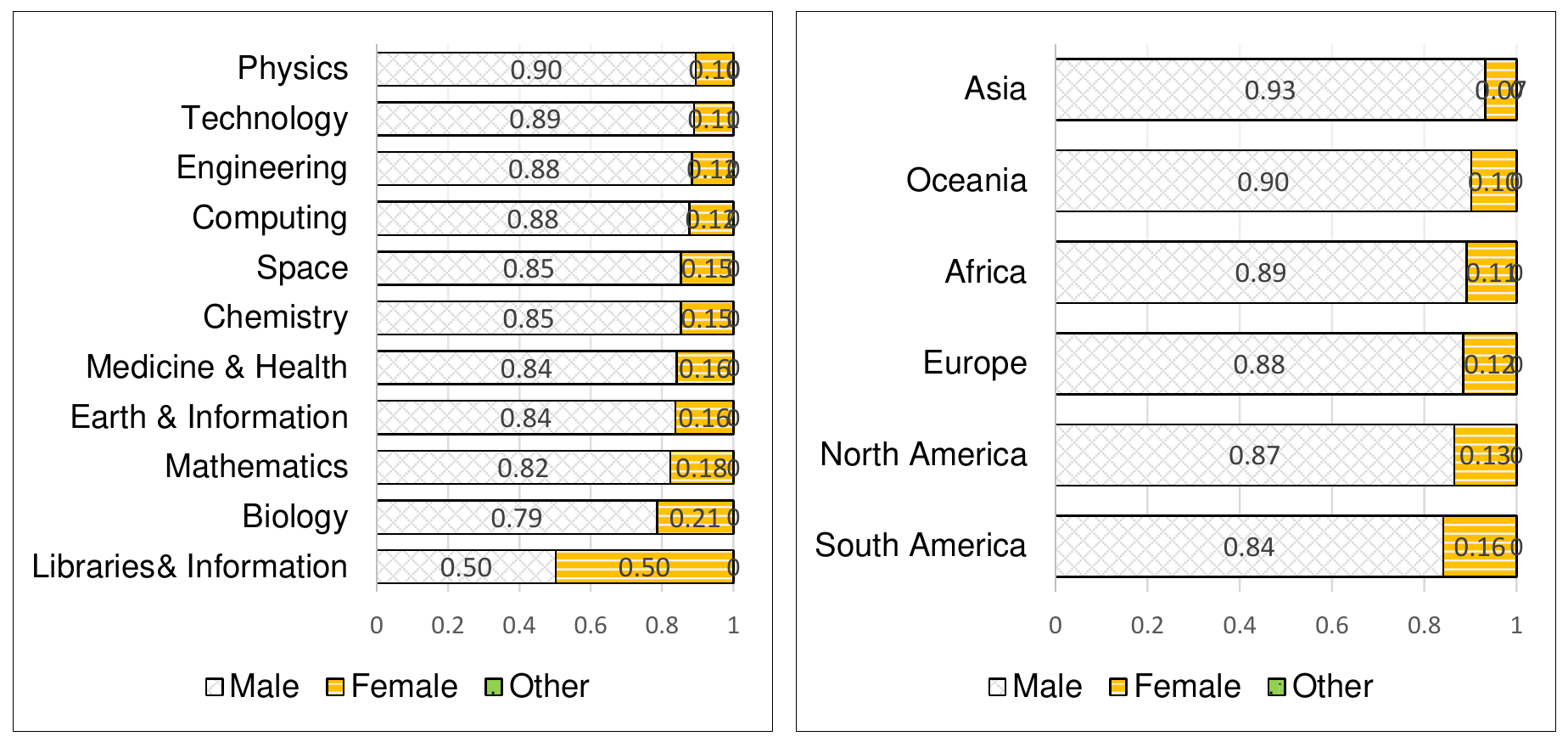}
\caption{Female award recipients in STEM (left) and female political office holders in different continents (right).} \label{fig:women}
\end{figure*}

\noindent
\textbf{Discovering Under-represented Groups.}
One use case for our data is in academic research in humanities. One standard social problem is identifying under-represented groups~\cite{ATKINSON2019423,doi:10.1144/SP506-2019-190,womeninpolitics}. These groups can be defined using one or more demographic factor, e.g., \textit{ethnicity}, \textit{gender}. Our data can be considered a resource to answer questions such as: is group X under-represented in community/domain Y? or what is the difference in representation of group X between different domains? A more specific example is shown in Figure~\ref{fig:women}. We show the fraction of \textit{female} award recipients (in Wikidata assigned \term{gender-female}) in STEM~\footnote{\href{https://en.wikipedia.org/wiki/Wikipedia:AfC\_sorting/STEM}{https://en.wikipedia.org/wiki/Wikipedia:AfC\_sorting/STEM}} and in political offices holders in different continents. These numbers can support needed initiatives for more representation of certain groups, e.g., programs such as \textit{Women in Tech}. In this example, we used Wiki-topics to specify spatial and topical dimensions, e.g., \textit{STEM.Physics} for Physics awards and an intersection of \textit{Regions.Europe and History\_and\_Society.Politics\_and\_government} for political offices in certain geographical regions.

Beyond the topics tool, our dataset is not isolated in terms of what we know about communities of interest and their members. Each subject or group can be linked to its Wikidata profile or Wikipedia article, allowing for more customized analyses. For instance, in award winning or holding public offices, statements are normally associated with temporal data in Wikidata, allowing the user of this dataset to explore progress across time. For instance, the charts in Figure~\ref{fig:women} can be re-plotted to include time windows, i.e., female award recipients in Physics [1960-1990], Physics [1991-present] and so on.

\noindent
\textbf{Exploring Cultural Differences in Governing.}
Our data can be used in political science research such as understanding governing in different parts of the world. One angle is to understand what kind of professions dominate public offices, e.g., presidents, mayors, governors, ministers, ... We compute these as communities of interest created using the property \term{position held}~\footnote{\href{https://www.wikidata.org/wiki/Property:P39}{https://www.wikidata.org/wiki/Property:P39}}. We retain communities of interest under \textit{Politics\_and\_Government}, and assign each to its equivalent geographical area, e.g., \textit{Central America} (see Table~\ref{tab:politicaloffice}). Note that we drop the profession at rank 1, since it is \textit{politician} for \textit{all} the areas, and that holding a certain office automatically turn one into a politician. This data can give better understanding of certain cultures or in case of democracies, how do people vote. \textit{Lawyer} is a recurring profession in Americas, especially in \textit{North America} with quarter of public office holders with \term{occupation-lawyer}.

\begin{table*}
\centering
\begin{tabular}{l|l}
\textbf{\cellcolor{gray!15}Area} & \textbf{\cellcolor{gray!15}Top professions}\\
Central Africa & diplomat {\small (0.27)}, economist {\small (0.04)}, civil servant {\small (0.01)}, philosopher {\small (0.01)}, minister {\small (0.01)}\\
Eastern Africa& diplomat {\small (0.09)}, judge {\small (0.03)}, lawyer {\small (0.03)}, military personnel {\small (0.03)}, economist {\small (0.02)}\\
Northern Africa & diplomat {\small (0.12)}, ruler {\small (0.12)}, lawyer {\small (0.03)}, military personnel {\small (0.01)}, imam {\small (0.01)}\\
Southern Africa& judge {\small (0.28)}, lawyer {\small (0.11)}, civil servant {\small (0.01)}, businessperson {\small (0.01)}\\
Western Africa& diplomat {\small (0.17)}, lawyer {\small (0.03)}, military personnel {\small (0.03)}, economist {\small (0.01)}, judge {\small (0.01)}\\
\hdashline
Central America& lawyer {\small (0.07)}, diplomat {\small (0.07)}, writer {\small (0.02)}, economist {\small (0.01)}, military personnel {\small (0.01)}\\
North America& lawyer {\small (0.25)}, diplomat {\small (0.06)}, judge {\small (0.03)}, military personnel {\small (0.01)}, businessperson {\small (0.01)}\\
South America& lawyer {\small (0.17)}, diplomat {\small (0.05)}, military personnel {\small (0.02)}, journalist {\small (0.01)}, historian {\small (0.01)}\\
\hdashline
East Asia& monarch {\small (0.09)}, diplomat {\small (0.07)}, lawyer {\small (0.06)}, judge {\small (0.06)}, prosecutor {\small (0.01)}\\
South Asia& diplomat {\small (0.05)}, lawyer {\small (0.03)}, economist {\small (0.02)}, civil servant {\small(0.02)}, judge {\small (0.01)}\\
Southeast Asia& sovereign {\small (0.09)}, judge {\small (0.08)}, lawyer {\small (0.07)}, military personnel {\small(0.03)}, diplomat {\small (0.02)}\\
West Asia& diplomat {\small (0.12)}, sovereign {\small (0.08)}, military personnel {\small (0.05)}, physician {\small (0.02)}, poet {\small (0.01)}\\
\hdashline
Eastern Europe& diplomat {\small (0.12)}, economist {\small (0.04)}, lawyer {\small (0.02)}, monarch {\small (0.02)}, university teacher {\small (0.01)}\\
Northern Europe& judge {\small (0.08)}, diplomat {\small (0.04)}, monarch {\small (0.02)}, lawyer {\small (0.02)}, journalist {\small (0.01)}\\
Southern Europe& diplomat {\small (0.07)}, lawyer {\small (0.04)}, military personnel {\small (0.02)}, jurist {\small (0.01)}, monarch {\small (0.01)}\\
Western Europe& lawyer {\small (0.13)}, judge {\small (0.06)}, diplomat {\small (0.03)}, military personnel {\small (0.02)}, suffragist {\small (0.02)}, teacher {\small (0.01)}\\
\hdashline
Oceania& lawyer {\small (0.08)}, diplomat {\small (0.04)},  judge  {\small (0.01)}, pastoralist {\small (0.01)}, solicitor {\small (0.01)}, farmer {\small (0.01)}\\
\end{tabular}
\caption{Top professions in political offices in different parts of the world.}
\label{tab:politicaloffice}
\end{table*}

\subsection{Edit Recommendations for Collaborative Encyclopedias}
The Web-scale collaborative knowledge base Wikidata contains more than 100 million items (or subjects) which have received almost 2 billion edits since its inception. Editors often need to prioritize their efforts, so useful tools to guide them can improve data quality and completeness, e.g., the Recoin plugin~\cite{10.1145/3184558.3191641} helps focus the editing on missing properties of subjects. Our approach and dataset can be used to improve this service by not only proposing relevant missing properties but also proposing a full statement about that property. As mentioned, we consider the PCA prior to inferring the negativity of a certain demographic factor. In that step, one cannot assert absent information but can offer a \textit{calculated guess} of what that might be, leaving it for human curators to confirm or deny. For example, \textit{Maja Vuković} is a member of winners of \textit{IBM fellowship}. For the property occupation in Wikidata, she has zero values~\footnote{\href{https://www.wikidata.org/wiki/Q111536437}{https://www.wikidata.org/wiki/Q111536437}} and Recoin lists \term{occupation} as the top missing property. Given the demographic data we have about professions of this community she is member of, we propose \textit{computer scientist, mathematician,} and \textit{engineer} as top 3 candidates. This can especially contribute to the completeness of information about long tail entities. Moreover, for subjects who are members of multiple Communities of Interest, one can merge similar demographic factors across communities and average confidence scores.

\subsection{Entity Summarization}
Web search results of queries about public figures can be improved by including salient and sometimes surprising facts. It increases user engagement~\cite{funfacts} to augment question answering and entity summarization results with \textit{did you know-like} statements. This is where our second dataset, i.e., the subject-centric,  shines. A user can use the autocompletion field in our web interface to query statements about subject of their choice. These are often surprising or unexpected statements. Examples:
\begin{itemize}
    \item Did you know that unlike 96\% of \textit{Oscar for Best Director winners}, \textit{Bong Joon-ho} does not speak \textit{English}, but \textit{Korean}?~\footnote{In fact he was accompanied by a translator to deliver his acceptance speech.}
    \item Did you know that unlike 88\% of winners of the \textit{Presidential Medal of Freedom}, Stephen Hawking is a non-American, but \textit{British}, winner?
\item Did you know that unlike 91\% of recipients of \textit{Liebig Medal} (established by Association of German Chemists), recipient \textit{Max Planck} was not a chemist but a physicist?
\end{itemize}

\section{Ethical \& FAIR Considerations}
In this work, datasets released are based on public information and tools provided by Wikimedia projects. We do not use any personal information. Every dataset record, i.e., JSON object, include all equivalent labels and IDs of properties and items retrieved from Wikidata (December 2022 version). Users can refer to these IDs for more details and definitions.

\section{Conclusion}
In this work, we generate demographics from Wiki-based sources by exploiting information about communities of interest. We release two datasets and publish a web interface for friendly browsing. Finally, we show three purposes for the data, namely as a resource for social sciences problems, edit recommender for collaborative knowledge bases, and salient fact generator for a better search experience. 

\section{Acknowledgments}
Funded by the German Research Foundation (DFG: Deutsche Forschungsgemeinschaft) - Project 453095897.

\bibliography{aaai22}

\end{document}